\documentclass[aps,prl,twocolumn,superscriptaddress]{revtex4-1}

\usepackage[colorlinks=true,citecolor=blue]{hyperref}
\usepackage{graphicx}
\usepackage{amsmath}
\usepackage{amssymb}

\begin{document}

\title{Moir\'e-enabled topological superconductivity}

\author{Shawulienu Kezilebieke}
\email{kezilebieke.a.shawulienu@jyu.fi}
\affiliation{Department of Applied Physics, Aalto University, 00076 Aalto, Finland}
\affiliation{Department of Physics, Department of Chemistry and Nanoscience Center, 
University of Jyväskylä, FI-40014 University of Jyväskylä, Finland}

\author{Viliam Va\v{n}o}
\affiliation{Department of Applied Physics, Aalto University, 00076 Aalto, Finland}

\author{Md N. Huda}
\affiliation{Department of Applied Physics, Aalto University, 00076 Aalto, Finland}

\author{Markus Aapro}
\affiliation{Department of Applied Physics, Aalto University, 00076 Aalto, Finland}

\author{Somesh C. Ganguli}
\affiliation{Department of Applied Physics, Aalto University, 00076 Aalto, Finland}

\author{Peter Liljeroth}
\affiliation{Department of Applied Physics, Aalto University, 00076 Aalto, Finland}

\author{Jose L. Lado}
\email{jose.lado@aalto.fi}
\affiliation{Department of Applied Physics, Aalto University, 00076 Aalto, Finland}



\begin{abstract}

The search for artificial topological superconductivity has been limited by the stringent conditions required for its emergence. As exemplified by the recent discoveries of various correlated electronic states in twisted van der Waals materials, moir\'e patterns can act as a powerful knob to create artificial electronic structures. Here, we demonstrate that a moir\'e pattern between a van der Waals superconductor and a monolayer ferromagnet creates a periodic potential modulation that enables the realization of a topological superconducting state that would not be accessible in the absence of the moir\'e. The magnetic moir\'e pattern gives rise to Yu-Shiba-Rusinov minibands and periodic modulation of the Majorana edge modes that we detect using low-temperature scanning tunneling microscopy (STM) and spectroscopy (STS). Moir\'e patterns and, more broadly, periodic potential modulations are powerful tools to overcome the conventional constrains for realizing and controlling topological superconductivity.
\end{abstract}

\maketitle

\vspace{21pt}

There are many routes to realizing topological superconductivity in artificial structures\cite{Beenakker2013,Alicea2012,RevModPhys.83.1057,Alicea2011,PhysRevX.6.031016,Nadj-Perge602,Feldman2016,PhysRevB.88.155420,PhysRevLett.114.236803,PhysRevB.96.174521}, and perhaps the most widely used path uses the combination of superconductivity, spin-orbit coupling and magnetism \cite{PhysRevLett.104.040502,PhysRevLett.105.177002}. This recipe has been recently used in a CrBr$_3$/NbSe$_2$ van der Waals (vdW) heterostructure, where the emergence of topological superconductivity was demonstrated \cite{Kezilebieke2020Nature}. In stark contrast with the realization based on semiconducting nanowires \cite{Mourik1003,Das2012,Deng1557}, the electronic structure of the CrBr$_3$/NbSe$_2$ system features a highly doped band of NbSe$_2$, far from the typical allowed regimes for topological superconductivity to appear. This system has a complex electronic structure combining a Fermi surface reconstruction of NbSe$_2$ stemming from its charge density wave \cite{Ugeda2015}, together with a strong moir\'e arising from the lattice mismatch between NbSe$_2$ and CrBr$_3$. It is surprising that such complex electronic structure with no external control parameters turns out to give rise to a state featuring topological superconductivity.

The emergence of topological superconductivity in a NbSe$_2$/CrBr$_3$ heterostructure\cite{Kezilebieke2020Nature} can be rationalized as follows. The ferromagnetic state of CrBr$_3$ induces an exchange field on the electronic structure of NbSe$_2$, and the mirror symmetry breaking of the heterostructure creates a strong Rashba spin-orbit coupling in the NbSe$_2$ bands. When including the intrinsic s-wave superconducting order of NbSe$_2$, the low energy electronic structure harvest the three fundamental ingredients for the emergence of artificial topological superconductivity\cite{Beenakker2013,Alicea2012,RevModPhys.83.1057,Alicea2011,PhysRevX.6.031016,Nadj-Perge602,Feldman2016,PhysRevB.88.155420,PhysRevLett.114.236803,PhysRevB.96.174521}: s-wave superconducting order, Rashba spin-orbit coupling and exchange fields. Such low-energy effective model has been shown to faithfully capture the phenomenology observed experimentally\cite{Kezilebieke2020Nature}. However, interesting additional microscopic contributions have been so far unaddressed. First, Majorana edge modes showed strong regular modulation at the edges of the topological superconducting island. Second, the mismatch between the NbSe$_2$ and CrBr$_3$ monolayers gives rise to a moir\'e pattern modulating all the parameters in space.  And finally, the emergence of topological superconductivity in the minimal model required a delicate fine-tuning of the NbSe$_2$ Fermi level. Here we extend our earlier experimental results on the CrBr$_3$/NbSe$_2$ system, demonstrating how the previous three features are naturally accounted by emergent moir\'e phenomena of the heterostructure. 

Here, we show that the apparent complexity created by the moir\'e pattern in the CrBr$_3$/NbSe$_2$ system can be the ultimate driving force of its topological superconducting state. In particular, the strongly modulated electrostatic potential and exchange coupling in the moir\'e heterostructure give rise to modulated 
Yu-Shiba-Rusinov (YSR) bands that allow for the emergence of topological superconductivity in generic regimes where it is otherwise forbidden. We explain theoretically and demonstrate experimentally that the moir\'e-modulation of the topological state emerging from the YSR bands is also visible in the spatial distribution of the one-dimensional topological Majorana modes. Our results put moir\'e physics forward as a powerful knob enabling topological superconductivity. Finally, conceptually similar effects can be realized by creating a periodic potential modulation (e.g.~through external gating) in semiconducting devices \cite{Mourik1003,Rokhinson2012,Das2012,PhysRevLett.110.176403,PhysRevLett.110.146404,PhysRevB.94.115128,PhysRevB.85.140508,PhysRevB.96.165147,Bttcher2018,PhysRevB.96.174521,PhysRevLett.125.086802}, which offers new ways of controlling topological superconductivity towards the realisation of topological qubits in the future.

\begin{figure*}[t!]
    \centering
    \includegraphics[width=.8\textwidth]{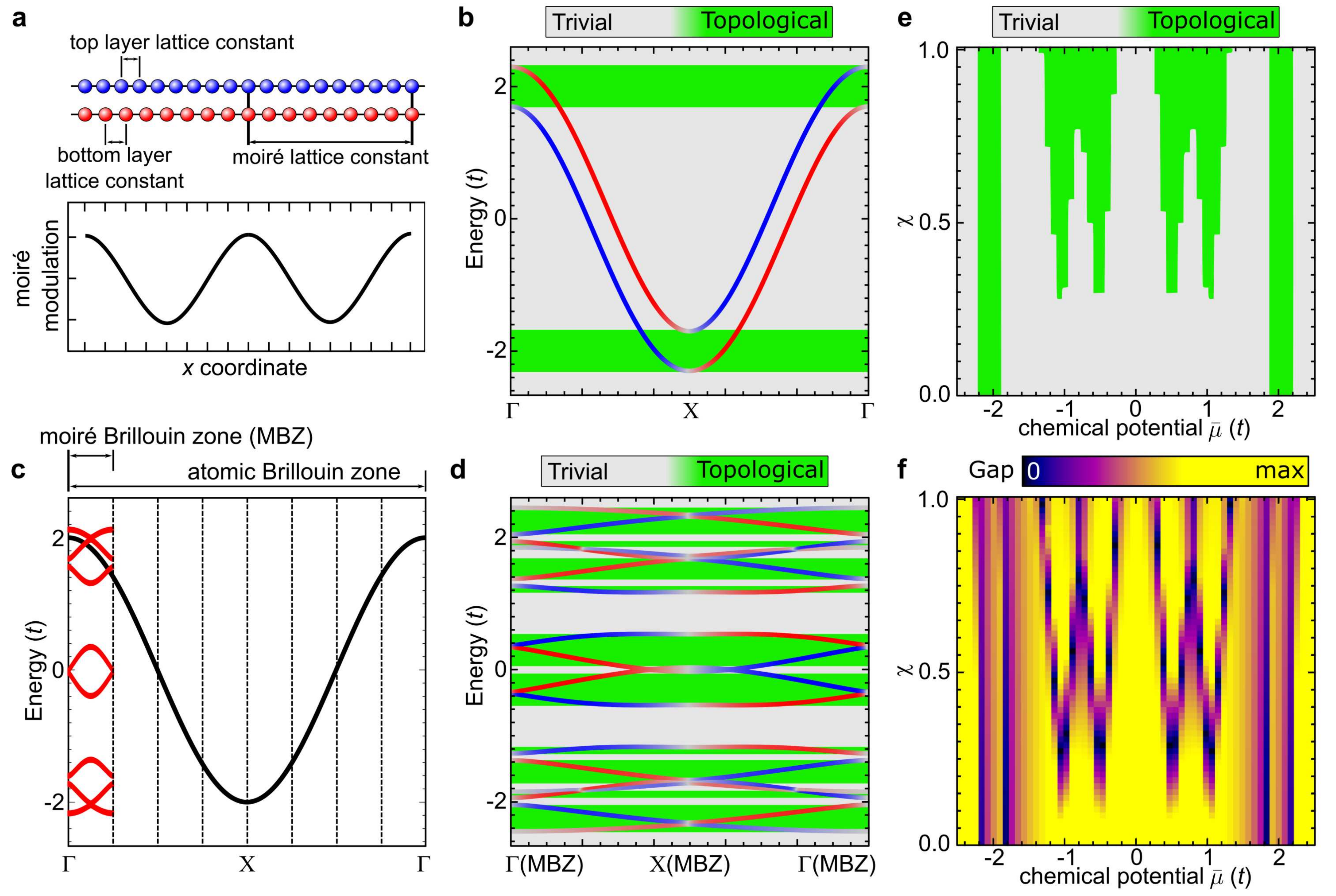}
	\caption{Moir\'e-driven topological superconductivity. (a) Schematic of the moir\'e pattern set up by the lattice mismatch, and the resulting modulation of the on-site energies, the exchange coupling and other parameters depending on the registry between the two lattices. (b) Pseudo-helical states appear in the presence of Rashba and exchange interactions at the top and bottom of the band of a one-dimensional model, giving rise to topological superconductivity (green). (c) Band structure of a simple one-dimensional model before (black line) and after (red lines) turning on the moir\'e modulation. (d) Including the effect of the moir\'e pattern enables pseudo-helical states and hence, topological superconductivity, to emerge at the top and bottom of the moir\'e mini-bands. Here, the moir\'e modulation is given by a modulation of the on-site energies. (e,f) Topological phase diagram (e) and the gap (f) for a 1D chain as function of the $\bar \mu$ and the the modulation $\chi$ of exchange.}
    \label{fig:sketch}
\end{figure*}

To understand the potential of moir\'e modulations for driving topological superconductivity, we consider a generic model incorporating  long-wavelength modulations in its different parameters \cite{Harper1955}. Specifically, we take a Hamiltonian that includes all the known ingredients for topological superconductivity: s-wave superconductivity, Rashba spin-orbit coupling, and ferromagnetism \cite{Beenakker2013,Alicea2012,RevModPhys.83.1057,PhysRevLett.105.077001,PhysRevLett.105.177002}. We introduce the moir\'e modulation through spatial variation of the parameters of the tight-binding model: on-site energies, the hoppings, the exchange coupling, Rashba spin-orbit coupling, and the s-wave superconductivity (details are given in the Supporting Information (SI)). Despite the increasing complexity of the Hamiltonian $H$ from having spatially dependent order parameters, their effects on enabling a topological superconducting state in arbitrary conditions can be easily rationalized. In order to illustrate these possibilities, we first focus on a minimal case: a one-dimensional moir\'e system (Fig.~\ref{fig:sketch}a).

For a one-dimensional model with uniform order parameters, topological superconductivity can only appear at the top bottom of the band, as shown in Fig.~\ref{fig:sketch}b. This is associated with a single set of pseudo-helical states that develop at the top and bottom of the band in the presence of Rashba spin-orbit interaction and exchange coupling. Turning on a moir\'e modulation in the chemical potential $\mu (\mathbf r) \sim \cos (\Omega x) $, for a given wavevector $\Omega$, will cause folding of the band structure and opening of minigaps between the folded bands as illustrated in Fig.~\ref{fig:sketch}c \cite{Harper1955,PhysRevLett.49.405,Bistritzer2011,PhysRevB.82.121407}. As shown in Fig.~\ref{fig:sketch}c, there are additional band tops and bottoms, where  topological superconductivity can potentially be realized. Indeed, when Rashba spin-orbit coupling, exchange and superconductivity are included in addition to the moir\'e pattern, pseudo-helical states appear close to charge neutrality (Fig.~\ref{fig:sketch}d), allowing for the emergence of topological superconductivity. This leads to topological regions in the phase diagram (Fig.~\ref{fig:sketch}e) at values of chemical potential corresponding to a topologically trivial state in the absence of the moir\'e modulation. Associated with new topological regions, gap closing and reopening is driven by the moir\'e modulation as shown in Fig.~\ref{fig:sketch}f. It is important to emphasize that in the absence of the moir\'e modulation, no topological superconducting state can be created at all in this energy range. While this example uses a modulation of the chemical potential, modulation in either exchange, Rashba, hoppings or proximity superconductivity are effective to drive these moir\'e-enabled topological phase transitions (Figure S1). This idea provides a new direction to explore topological superconductivity in designed one-dimensional systems, such as nanowires grown with a long-range modulation \cite{Algra2008,Zhang2019}, in doping regimes in which it would not be allowed otherwise.

\begin{figure}[t!]
    \centering
    \includegraphics[width=\columnwidth]{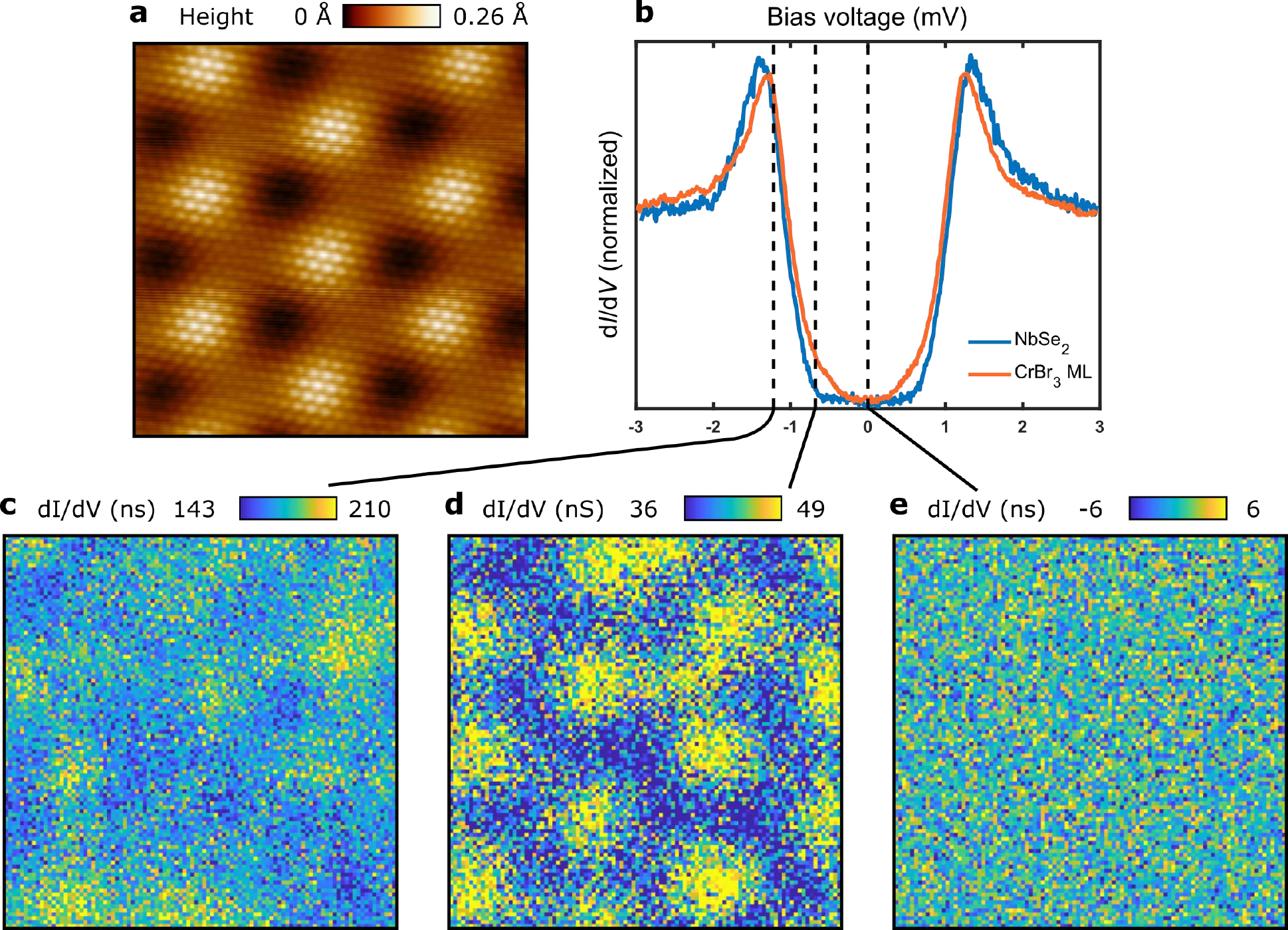}
	\caption{Correlation between moir\'e pattern and YSR bands. (a) STM image of CrBr$_3$ ML on NbSe$_2$ obtained at $V = 1.5$ V and $I$ = $300$ pA, image size is $20 \times 20$ nm$^2$. (b) d$I$/d$V$ spectroscopy on the NbSe$_2$ substrate (blue) and on the island of CrBr$_3$ ML (orange). (c-e) d$I$/d$V$ maps (from spectral grid) at different $V$ values highlighted in (b). The maps correspond to the STM image in (a) and each map is plotted on a separate color scale. A clear correlation between the moir\'e pattern and d$I$/d$V$ map only appears at the energy of the YSR bands ($V = 0.8$ mV). The grid map measurement conditions: V = 3 mV, I = 300 pA, and T = 350 mK}
    \label{Fig2_real}
\end{figure}

This phenomenology can be extended to two-dimensional systems that naturally arise due to the moir\'e modulation in van der Waals heterostructures. Fig.~\ref{Fig2_real}a shows an atomically resolved STM image of the CrBr$_3$ monolayer grown on a bulk NbSe$_2$ substrate (see SI for experimental details), revealing a well-ordered moir\'e superstructure with 6.3 nm periodicity arising from the lattice mismatch between the CrBr$_3$ and the NbSe$_2$ layers \cite{Kezilebieke2020}.
The moir\'e pattern matches a structure with 19 NbSe$_2$ unit cells accommodating 10 unit cells of CrBr$_3$, thus forming a 6.3 nm $\times$ 6.3 nm superstructure. This also matches the measured lattice constants of CrBr$_3$ and NbSe$_2$.

The interaction of the magnetism of the CrBr$_3$ layer \cite{Chen2019,Kezilebieke2020Nature,Kezilebieke2020} with the superconductivity from the NbSe$_2$ substrate gives rise to the YSR bands inside the superconducting gap that are also modulated by the moir\'e pattern. The formation of YSR band is shown in Fig.~\ref{Fig2_real}b (orange line) where the d$I$/d$V$ spectrum taken in the middle of the CrBr$_3$ island has a pair of conductance onsets at $\pm 0.35$ mV. This spectroscopic signature can be compared to a d$I$/d$V$ spectrum of bare NbSe$_2$, where a hard gap with an extended region of zero differential conductance around zero bias is observed (Fig.~\ref{Fig2_real}b, blue line). By subtracting the background spectra from the two-band model fit, we obtain the experimental topological gap that is around $\Delta_t\approx0.3\Delta$ (see SI). 
In order to visualize the spatial modulation of the YSR band, we have recorded grid d$I$/d$V$ spectroscopy maps (Fig.~\ref{Fig2_real}c-e) over the area shown in the Fig.~\ref{Fig2_real}a. The d$I$/d$V$ maps exhibit periodic modulation of the signal intensity over the moir\'e unit cell only at the energy of the YSR bands. This is caused by the intensity variations of the YSR band local density of states (LDOS), rather than energy variations of the YSR band as further demonstrated in the SI (Figure S7, S8).

\begin{figure}[t!]
    \centering
    \includegraphics[width=\columnwidth]{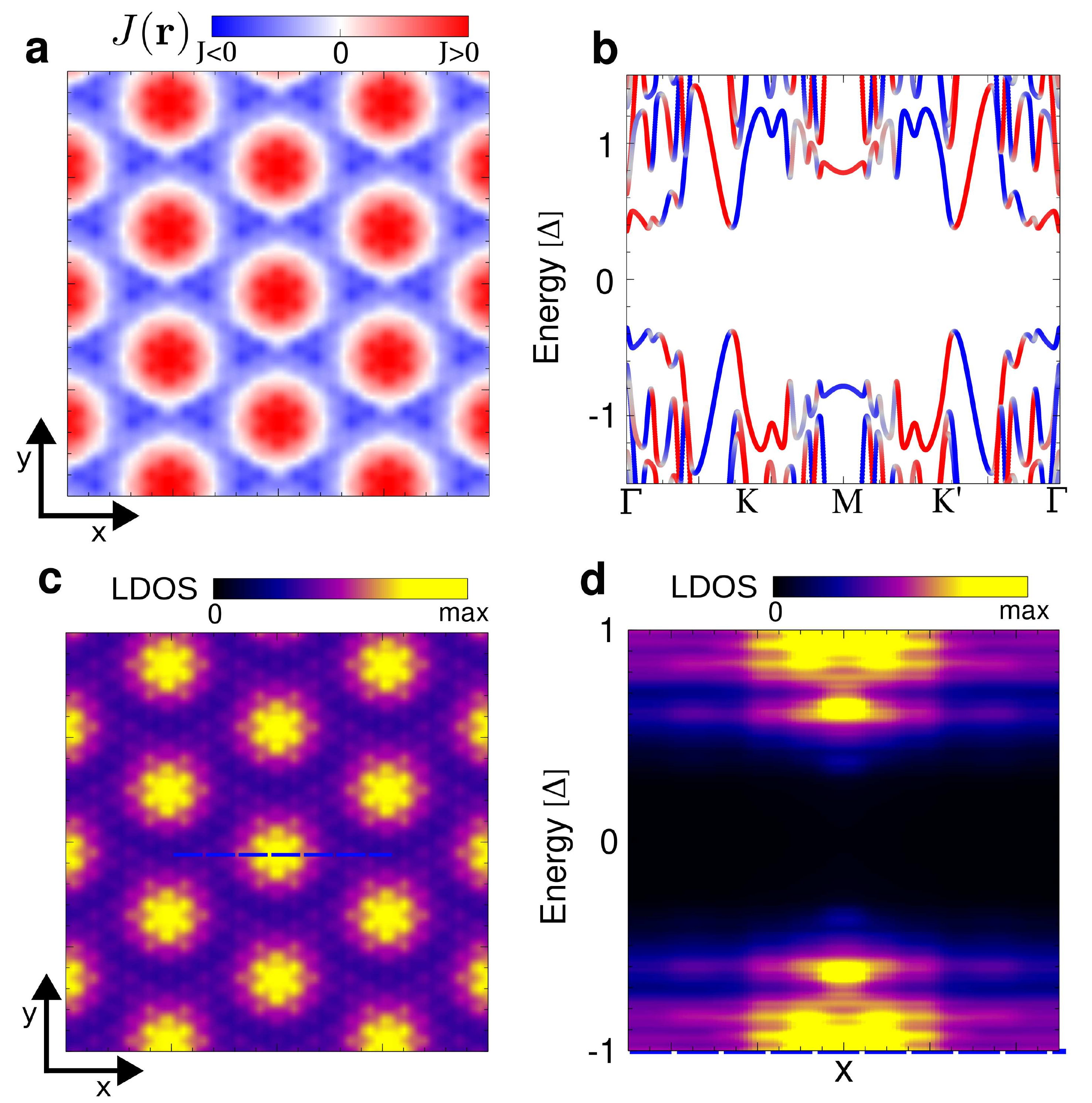}
	\caption{Moir\'e-enabled topological superconductivity: mechanism in 2D topological superconductor NbSe$_2$/CrBr$_3$. (a) Moir\'e profile for the exchange coupling $J(\mathbf r)$. (b) Resulting band structure in the topological regime with spatially varying exchange as described in the text. (c) The modulations of the YSR band LDOS due to the modulated exchange. (d) Spatial modulation of the bulk YSR bands along the line indicated in panel (c).	}
    \label{fig:theo}
\end{figure}

\begin{figure*}[t!]
    \centering
    \includegraphics[width=0.7\textwidth]{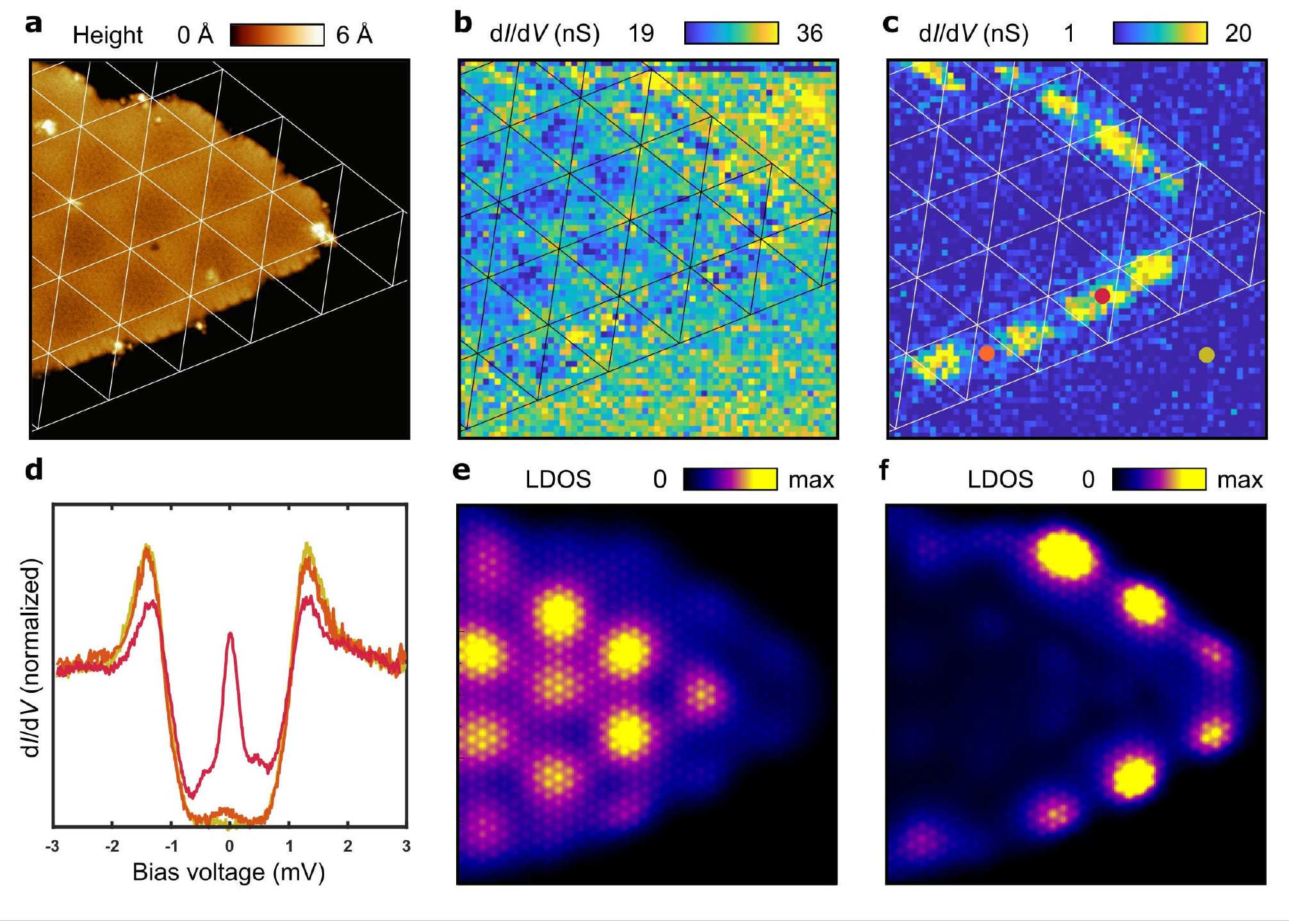}
	\caption{Edge states of the topological superconductor in van der Waals heterostructure with a moir\'e pattern. (a) 33 $\times$ 33 nm$^2$ STM image of CrBr$_3$ island on NbSe$_2$, obtained at $V = 1.0$ V and $I$ = 10 pA. (b) d$I$/d$V$ map at energy of the YSR bands $V = 0.8$ mV with a moir\'e pattern appearing in the bulk of CrBr$_3$ island. (c) d$I$/d$V$ map (from spectral grid) at $V = 0$ mV showing Majorana edge modes. (d) d$I$/d$V$ spectra acquired at the positions indicated in panel (c) (red - strong edge mode intensity, orange - weak edge mode intensity, and yellow - background spectrum on NbSe$_2$. (e,f) Theorically computed LDOS in the presence of a moir\'e exchange, at the energies of the YSR states (e) and the Majorana zero modes (f) for an island that closely mimics the shape of the studied experimentally. The grid map measurement conditions: V = 3 mV, I = 140 pA, and T = 350 mK}
    \label{fig:edge}
\end{figure*}

The microscopic origin of the variations of the YSR band intensities can be easily rationalized. First, the modulation of exchange (Fig.~\ref{fig:theo}a) stems from the strong dependence of superexchange interactions on the local stacking, as demonstrated in CrI$_3$ and CrBr$_3$ bilayers \cite{Sivadas2018,Song2019,Zhang2019}. This feature suggests that the moir\'e pattern not only modulates the absolute value of the effective exchange, but it also can change its sign \cite{Sivadas2018,Song2019}. Second, as a consequence of the modulation of the exchange field, the superconducting order parameter will also be modulated in the opposite way, due to the competition of s-wave superconductivity from NbSe$_2$ and the proximity induced exchange field \cite{PhysRevX.10.041003}. Third, the modulation of the onsite energies stems from an electrostatic effect associated with the stacking. We can directly measure the modulation of this electrostatic potential through the spatial modulation of the conduction band edge of CrBr$_3$ (Figure S9). Fourth, modulations in the hoppings are expected from small relaxation effects, well-known in other dichalcogenide-based twisted systems \cite{PhysRevLett.121.266401,PhysRevLett.124.206101}. Fifth, the charge density wave of NbSe$_2$\cite{Ugeda2015} will introduce additional short-range modulation both in the hopping and local onsite energies \cite{PhysRevB.97.081101,Guster2019}. While all these effects can be incorporated into the effective model, we can reproduce the experimental results even using a minimal model that only incorporates spatially varying exchange interactions and onsite energies (Fig.~\ref{fig:theo}a). This is supported by the fact that a simple triangular lattice nearest neighbor tight binding model gives a good representation of the Fermi surface of NbSe$_2$ in the presence of Ising and Rashba SOCs and the CDW reconstruction (see SI Section ``Realistic tight-binding model for CrBr$_3$/NbSe$_2$ heterostructure'' for details). This results in a topological superconducting band structure (Fig.~\ref{fig:theo}b) in a chemical potential range close to charge neutrality, where the system is trivial in the absence of moir\'e modulations. Associated with these modulations, moir\'e-modulated YSR bands emerge (Fig.~\ref{fig:theo}c). The theory predicts - in agreement with our experimental results - that the moir\'e pattern gives rise to a spatial modulation of the intensity (Fig.~\ref{fig:theo}d) of the in-gap states but not of their energies.

Above, we have focused on the impact of the moir\'e pattern on the bulk electronic structure, but the the moir\'e electronic structure also gets imprinted on the topological edge modes. In particular, the emergence of a YSR moir\'e band structure suggests that topological edge modes may inherit the moir\'e distribution of the bulk YSR states. We now focus on the edge of a CrBr$_3$ island, as shown in Fig.~\ref{fig:edge}a. At biases at and above the YSR bands (Fig.~\ref{fig:edge}b), no strong modulation at the edge is observed. In stark contrast, when taking energies inside the topological gap, we observe topological edge modes with a strong modulation with the period of the moir\'e pattern (Fig.~\ref{fig:edge}c). This is also visible in the single d$I$/d$V$ spectra extracted at the points corresponding to the minimum and maximum intensity along the edge (points marked in Fig.~\ref{fig:edge}c, spectra shown in Fig.~\ref{fig:edge}d). We have analyzed a corresponding finite-size structure with our theoretical model (details in the SI) as shown in Figs.~\ref{fig:edge}e,f. As expected, at energies above the topological gap, modulated YSR states appear (Fig.~\ref{fig:edge}e). In strong correspondence with the experimental results, inside the gap, strongly modulated in-gap modes dominate the spectra (Fig.~\ref{fig:edge}f). This direct relationship between the edge modes and the bulk moir\'e modulation demonstrates a non-trivial role of the moir\'e pattern in creating the topological superconducting state. The moir\'e-induced topological phase transition and the modulation of edge modes are general features of the physical picture and will occur for the one-dimensional and two-dimensional realizations of these systems. This provides an experimentally simple way of verifying the presence and assessing the impact of the moir\'e modulation on the topological superconducting state.

To summarize, we have demonstrated that moir\'e modulations allow realization of topological superconductivity in parameter regimes otherwise forbidden by the electronic structure. 
In particular, by accounting for the moire modulation, we have solved three open questions on the
emergence of topological superconductivity in CrBr$_3$/NbSe$_2$.
First, the spatial modulation of the edge modes directly corresponds to the moir\'e modulation of the bulk Yu-Shiba-Rusinov bands. Second, there is no need for the fine tuning of the chemical potential as the moir\'e modulation results in a topological phase around charge neutrality over a broad range of the values of the chemical potential. 
And thirs, the detrimental effect of the Ising spin-orbit interaction is mostly removed by the modification of the band structure due to the charge-density wave modulation of the NbSe$_2$.
Concomitant to this moir\'e-enabled superconducting state, moir\'e-modulated
YSR bands appear, whose topological band structure is ultimately responsible
for the topological superconducting state. We have demonstrated this idea in a
    CrBr$_3$/NbSe$_2$ twisted heterostructure, showing the emergence of moir\'e
    YSR bands and moir\'e-modulated edge modes, the two paradigmatic
    experimental signatures of a moir\'e-enabled topological state.
    Moir\'e-enabled topological phase transitions are especially powerful in
    twisted van der Waals heterostructures, where the twist angle can be used
    as a knob to push the system to a topological superconducting state. Our
    results demonstrate the possibility of using twist engineering to design
    topological quantum materials with a high potential for creating a platform
    for realizing strongly interacting topological superconductors. This
        provides a new paradigmatic direction in the field of topological
        twistronics.


\textbf{Acknowledgements}: This research made use of the Aalto Nanomicroscopy Center (Aalto NMC) facilities and was supported by the European Research Council (ERC-2017-AdG no.~788185 ``Artificial Designer Materials''), Academy of Finland (Academy professor funding nos.~318995 and 320555, Academy research fellow nos.~331342, 336243 and no.~338478 and 346654), and the Jane and Aatos Erkko Foundation. We acknowledge the computational resources provided by the Aalto Science-IT project.

\bibliography{biblio}

\begin{thebibliography}{41}%
\makeatletter
\providecommand \@ifxundefined [1]{%
 \@ifx{#1\undefined}
}%
\providecommand \@ifnum [1]{%
 \ifnum #1\expandafter \@firstoftwo
 \else \expandafter \@secondoftwo
 \fi
}%
\providecommand \@ifx [1]{%
 \ifx #1\expandafter \@firstoftwo
 \else \expandafter \@secondoftwo
 \fi
}%
\providecommand \natexlab [1]{#1}%
\providecommand \enquote  [1]{``#1''}%
\providecommand \bibnamefont  [1]{#1}%
\providecommand \bibfnamefont [1]{#1}%
\providecommand \citenamefont [1]{#1}%
\providecommand \href@noop [0]{\@secondoftwo}%
\providecommand \href [0]{\begingroup \@sanitize@url \@href}%
\providecommand \@href[1]{\@@startlink{#1}\@@href}%
\providecommand \@@href[1]{\endgroup#1\@@endlink}%
\providecommand \@sanitize@url [0]{\catcode `\\12\catcode `\$12\catcode
  `\&12\catcode `\#12\catcode `\^12\catcode `\_12\catcode `\%12\relax}%
\providecommand \@@startlink[1]{}%
\providecommand \@@endlink[0]{}%
\providecommand \url  [0]{\begingroup\@sanitize@url \@url }%
\providecommand \@url [1]{\endgroup\@href {#1}{\urlprefix }}%
\providecommand \urlprefix  [0]{URL }%
\providecommand \Eprint [0]{\href }%
\providecommand \doibase [0]{http://dx.doi.org/}%
\providecommand \selectlanguage [0]{\@gobble}%
\providecommand \bibinfo  [0]{\@secondoftwo}%
\providecommand \bibfield  [0]{\@secondoftwo}%
\providecommand \translation [1]{[#1]}%
\providecommand \BibitemOpen [0]{}%
\providecommand \bibitemStop [0]{}%
\providecommand \bibitemNoStop [0]{.\EOS\space}%
\providecommand \EOS [0]{\spacefactor3000\relax}%
\providecommand \BibitemShut  [1]{\csname bibitem#1\endcsname}%
\let\auto@bib@innerbib\@empty
\bibitem [{\citenamefont {Beenakker}(2013)}]{Beenakker2013}%
  \BibitemOpen
  \bibfield  {author} {\bibinfo {author} {\bibfnamefont {C.}~\bibnamefont
  {Beenakker}},\ }\href {\doibase 10.1146/annurev-conmatphys-030212-184337}
  {\bibfield  {journal} {\bibinfo  {journal} {Annu. Rev. Condens. Matter
  Phys.}\ }\textbf {\bibinfo {volume} {4}},\ \bibinfo {pages} {113} (\bibinfo
  {year} {2013})}\BibitemShut {NoStop}%
\bibitem [{\citenamefont {Alicea}(2012)}]{Alicea2012}%
  \BibitemOpen
  \bibfield  {author} {\bibinfo {author} {\bibfnamefont {J.}~\bibnamefont
  {Alicea}},\ }\href {\doibase 10.1088/0034-4885/75/7/076501} {\bibfield
  {journal} {\bibinfo  {journal} {Rep. Prog. Phys.}\ }\textbf {\bibinfo
  {volume} {75}},\ \bibinfo {pages} {076501} (\bibinfo {year}
  {2012})}\BibitemShut {NoStop}%
\bibitem [{\citenamefont {Qi}\ and\ \citenamefont
  {Zhang}(2011)}]{RevModPhys.83.1057}%
  \BibitemOpen
  \bibfield  {author} {\bibinfo {author} {\bibfnamefont {X.-L.}\ \bibnamefont
  {Qi}}\ and\ \bibinfo {author} {\bibfnamefont {S.-C.}\ \bibnamefont {Zhang}},\
  }\href {\doibase 10.1103/RevModPhys.83.1057} {\bibfield  {journal} {\bibinfo
  {journal} {Rev. Mod. Phys.}\ }\textbf {\bibinfo {volume} {83}},\ \bibinfo
  {pages} {1057} (\bibinfo {year} {2011})}\BibitemShut {NoStop}%
\bibitem [{\citenamefont {Alicea}\ \emph {et~al.}(2011)\citenamefont {Alicea},
  \citenamefont {Oreg}, \citenamefont {Refael}, \citenamefont {von Oppen},\
  and\ \citenamefont {Fisher}}]{Alicea2011}%
  \BibitemOpen
  \bibfield  {author} {\bibinfo {author} {\bibfnamefont {J.}~\bibnamefont
  {Alicea}}, \bibinfo {author} {\bibfnamefont {Y.}~\bibnamefont {Oreg}},
  \bibinfo {author} {\bibfnamefont {G.}~\bibnamefont {Refael}}, \bibinfo
  {author} {\bibfnamefont {F.}~\bibnamefont {von Oppen}}, \ and\ \bibinfo
  {author} {\bibfnamefont {M.~P.~A.}\ \bibnamefont {Fisher}},\ }\href {\doibase
  10.1038/nphys1915} {\bibfield  {journal} {\bibinfo  {journal} {Nat. Phys.}\
  }\textbf {\bibinfo {volume} {7}},\ \bibinfo {pages} {412} (\bibinfo {year}
  {2011})}\BibitemShut {NoStop}%
\bibitem [{\citenamefont {Aasen}\ \emph {et~al.}(2016)\citenamefont {Aasen},
  \citenamefont {Hell}, \citenamefont {Mishmash}, \citenamefont {Higginbotham},
  \citenamefont {Danon}, \citenamefont {Leijnse}, \citenamefont {Jespersen},
  \citenamefont {Folk}, \citenamefont {Marcus}, \citenamefont {Flensberg},\
  and\ \citenamefont {Alicea}}]{PhysRevX.6.031016}%
  \BibitemOpen
  \bibfield  {author} {\bibinfo {author} {\bibfnamefont {D.}~\bibnamefont
  {Aasen}}, \bibinfo {author} {\bibfnamefont {M.}~\bibnamefont {Hell}},
  \bibinfo {author} {\bibfnamefont {R.~V.}\ \bibnamefont {Mishmash}}, \bibinfo
  {author} {\bibfnamefont {A.}~\bibnamefont {Higginbotham}}, \bibinfo {author}
  {\bibfnamefont {J.}~\bibnamefont {Danon}}, \bibinfo {author} {\bibfnamefont
  {M.}~\bibnamefont {Leijnse}}, \bibinfo {author} {\bibfnamefont {T.~S.}\
  \bibnamefont {Jespersen}}, \bibinfo {author} {\bibfnamefont {J.~A.}\
  \bibnamefont {Folk}}, \bibinfo {author} {\bibfnamefont {C.~M.}\ \bibnamefont
  {Marcus}}, \bibinfo {author} {\bibfnamefont {K.}~\bibnamefont {Flensberg}}, \
  and\ \bibinfo {author} {\bibfnamefont {J.}~\bibnamefont {Alicea}},\ }\href
  {\doibase 10.1103/PhysRevX.6.031016} {\bibfield  {journal} {\bibinfo
  {journal} {Phys. Rev. X}\ }\textbf {\bibinfo {volume} {6}},\ \bibinfo {pages}
  {031016} (\bibinfo {year} {2016})}\BibitemShut {NoStop}%
\bibitem [{\citenamefont {Nadj-Perge}\ \emph {et~al.}(2014)\citenamefont
  {Nadj-Perge}, \citenamefont {Drozdov}, \citenamefont {Li}, \citenamefont
  {Chen}, \citenamefont {Jeon}, \citenamefont {Seo}, \citenamefont {MacDonald},
  \citenamefont {Bernevig},\ and\ \citenamefont {Yazdani}}]{Nadj-Perge602}%
  \BibitemOpen
  \bibfield  {author} {\bibinfo {author} {\bibfnamefont {S.}~\bibnamefont
  {Nadj-Perge}}, \bibinfo {author} {\bibfnamefont {I.~K.}\ \bibnamefont
  {Drozdov}}, \bibinfo {author} {\bibfnamefont {J.}~\bibnamefont {Li}},
  \bibinfo {author} {\bibfnamefont {H.}~\bibnamefont {Chen}}, \bibinfo {author}
  {\bibfnamefont {S.}~\bibnamefont {Jeon}}, \bibinfo {author} {\bibfnamefont
  {J.}~\bibnamefont {Seo}}, \bibinfo {author} {\bibfnamefont {A.~H.}\
  \bibnamefont {MacDonald}}, \bibinfo {author} {\bibfnamefont {B.~A.}\
  \bibnamefont {Bernevig}}, \ and\ \bibinfo {author} {\bibfnamefont
  {A.}~\bibnamefont {Yazdani}},\ }\href {\doibase 10.1126/science.1259327}
  {\bibfield  {journal} {\bibinfo  {journal} {Science}\ }\textbf {\bibinfo
  {volume} {346}},\ \bibinfo {pages} {602} (\bibinfo {year} {2014})},\ \Eprint
  {http://arxiv.org/abs/https://science.sciencemag.org/content/346/6209/602.full.pdf}
  {https://science.sciencemag.org/content/346/6209/602.full.pdf} \BibitemShut
  {NoStop}%
\bibitem [{\citenamefont {Feldman}\ \emph {et~al.}(2016)\citenamefont
  {Feldman}, \citenamefont {Randeria}, \citenamefont {Li}, \citenamefont
  {Jeon}, \citenamefont {Xie}, \citenamefont {Wang}, \citenamefont {Drozdov},
  \citenamefont {Bernevig},\ and\ \citenamefont {Yazdani}}]{Feldman2016}%
  \BibitemOpen
  \bibfield  {author} {\bibinfo {author} {\bibfnamefont {B.~E.}\ \bibnamefont
  {Feldman}}, \bibinfo {author} {\bibfnamefont {M.~T.}\ \bibnamefont
  {Randeria}}, \bibinfo {author} {\bibfnamefont {J.}~\bibnamefont {Li}},
  \bibinfo {author} {\bibfnamefont {S.}~\bibnamefont {Jeon}}, \bibinfo {author}
  {\bibfnamefont {Y.}~\bibnamefont {Xie}}, \bibinfo {author} {\bibfnamefont
  {Z.}~\bibnamefont {Wang}}, \bibinfo {author} {\bibfnamefont {I.~K.}\
  \bibnamefont {Drozdov}}, \bibinfo {author} {\bibfnamefont {B.~A.}\
  \bibnamefont {Bernevig}}, \ and\ \bibinfo {author} {\bibfnamefont
  {A.}~\bibnamefont {Yazdani}},\ }\href {\doibase 10.1038/nphys3947} {\bibfield
   {journal} {\bibinfo  {journal} {Nat. Phys.}\ }\textbf {\bibinfo {volume}
  {13}},\ \bibinfo {pages} {286} (\bibinfo {year} {2016})}\BibitemShut
  {NoStop}%
\bibitem [{\citenamefont {Pientka}\ \emph {et~al.}(2013)\citenamefont
  {Pientka}, \citenamefont {Glazman},\ and\ \citenamefont {von
  Oppen}}]{PhysRevB.88.155420}%
  \BibitemOpen
  \bibfield  {author} {\bibinfo {author} {\bibfnamefont {F.}~\bibnamefont
  {Pientka}}, \bibinfo {author} {\bibfnamefont {L.~I.}\ \bibnamefont
  {Glazman}}, \ and\ \bibinfo {author} {\bibfnamefont {F.}~\bibnamefont {von
  Oppen}},\ }\href {\doibase 10.1103/PhysRevB.88.155420} {\bibfield  {journal}
  {\bibinfo  {journal} {Phys. Rev. B}\ }\textbf {\bibinfo {volume} {88}},\
  \bibinfo {pages} {155420} (\bibinfo {year} {2013})}\BibitemShut {NoStop}%
\bibitem [{\citenamefont {R\"ontynen}\ and\ \citenamefont
  {Ojanen}(2015)}]{PhysRevLett.114.236803}%
  \BibitemOpen
  \bibfield  {author} {\bibinfo {author} {\bibfnamefont {J.}~\bibnamefont
  {R\"ontynen}}\ and\ \bibinfo {author} {\bibfnamefont {T.}~\bibnamefont
  {Ojanen}},\ }\href {\doibase 10.1103/PhysRevLett.114.236803} {\bibfield
  {journal} {\bibinfo  {journal} {Phys. Rev. Lett.}\ }\textbf {\bibinfo
  {volume} {114}},\ \bibinfo {pages} {236803} (\bibinfo {year}
  {2015})}\BibitemShut {NoStop}%
\bibitem [{\citenamefont {P\"oyh\"onen}\ and\ \citenamefont
  {Ojanen}(2017)}]{PhysRevB.96.174521}%
  \BibitemOpen
  \bibfield  {author} {\bibinfo {author} {\bibfnamefont {K.}~\bibnamefont
  {P\"oyh\"onen}}\ and\ \bibinfo {author} {\bibfnamefont {T.}~\bibnamefont
  {Ojanen}},\ }\href {\doibase 10.1103/PhysRevB.96.174521} {\bibfield
  {journal} {\bibinfo  {journal} {Phys. Rev. B}\ }\textbf {\bibinfo {volume}
  {96}},\ \bibinfo {pages} {174521} (\bibinfo {year} {2017})}\BibitemShut
  {NoStop}%
\bibitem [{\citenamefont {Sau}\ \emph {et~al.}(2010)\citenamefont {Sau},
  \citenamefont {Lutchyn}, \citenamefont {Tewari},\ and\ \citenamefont
  {Das~Sarma}}]{PhysRevLett.104.040502}%
  \BibitemOpen
  \bibfield  {author} {\bibinfo {author} {\bibfnamefont {J.~D.}\ \bibnamefont
  {Sau}}, \bibinfo {author} {\bibfnamefont {R.~M.}\ \bibnamefont {Lutchyn}},
  \bibinfo {author} {\bibfnamefont {S.}~\bibnamefont {Tewari}}, \ and\ \bibinfo
  {author} {\bibfnamefont {S.}~\bibnamefont {Das~Sarma}},\ }\href {\doibase
  10.1103/PhysRevLett.104.040502} {\bibfield  {journal} {\bibinfo  {journal}
  {Phys. Rev. Lett.}\ }\textbf {\bibinfo {volume} {104}},\ \bibinfo {pages}
  {040502} (\bibinfo {year} {2010})}\BibitemShut {NoStop}%
\bibitem [{\citenamefont {Oreg}\ \emph {et~al.}(2010)\citenamefont {Oreg},
  \citenamefont {Refael},\ and\ \citenamefont {von
  Oppen}}]{PhysRevLett.105.177002}%
  \BibitemOpen
  \bibfield  {author} {\bibinfo {author} {\bibfnamefont {Y.}~\bibnamefont
  {Oreg}}, \bibinfo {author} {\bibfnamefont {G.}~\bibnamefont {Refael}}, \ and\
  \bibinfo {author} {\bibfnamefont {F.}~\bibnamefont {von Oppen}},\ }\href
  {\doibase 10.1103/PhysRevLett.105.177002} {\bibfield  {journal} {\bibinfo
  {journal} {Phys. Rev. Lett.}\ }\textbf {\bibinfo {volume} {105}},\ \bibinfo
  {pages} {177002} (\bibinfo {year} {2010})}\BibitemShut {NoStop}%
\bibitem [{\citenamefont {Kezilebieke}\ \emph {et~al.}(2020)\citenamefont
  {Kezilebieke}, \citenamefont {Huda}, \citenamefont {Va{\v{n}}o},
  \citenamefont {Aapro}, \citenamefont {Ganguli}, \citenamefont {Silveira},
  \citenamefont {G{\l}odzik}, \citenamefont {Foster}, \citenamefont {Ojanen},\
  and\ \citenamefont {Liljeroth}}]{Kezilebieke2020Nature}%
  \BibitemOpen
  \bibfield  {author} {\bibinfo {author} {\bibfnamefont {S.}~\bibnamefont
  {Kezilebieke}}, \bibinfo {author} {\bibfnamefont {M.~N.}\ \bibnamefont
  {Huda}}, \bibinfo {author} {\bibfnamefont {V.}~\bibnamefont {Va{\v{n}}o}},
  \bibinfo {author} {\bibfnamefont {M.}~\bibnamefont {Aapro}}, \bibinfo
  {author} {\bibfnamefont {S.~C.}\ \bibnamefont {Ganguli}}, \bibinfo {author}
  {\bibfnamefont {O.~J.}\ \bibnamefont {Silveira}}, \bibinfo {author}
  {\bibfnamefont {S.}~\bibnamefont {G{\l}odzik}}, \bibinfo {author}
  {\bibfnamefont {A.~S.}\ \bibnamefont {Foster}}, \bibinfo {author}
  {\bibfnamefont {T.}~\bibnamefont {Ojanen}}, \ and\ \bibinfo {author}
  {\bibfnamefont {P.}~\bibnamefont {Liljeroth}},\ }\href {\doibase
  10.1038/s41586-020-2989-y} {\bibfield  {journal} {\bibinfo  {journal}
  {Nature}\ }\textbf {\bibinfo {volume} {588}},\ \bibinfo {pages} {424}
  (\bibinfo {year} {2020})}\BibitemShut {NoStop}%
\bibitem [{\citenamefont {Mourik}\ \emph {et~al.}(2012)\citenamefont {Mourik},
  \citenamefont {Zuo}, \citenamefont {Frolov}, \citenamefont {Plissard},
  \citenamefont {Bakkers},\ and\ \citenamefont {Kouwenhoven}}]{Mourik1003}%
  \BibitemOpen
  \bibfield  {author} {\bibinfo {author} {\bibfnamefont {V.}~\bibnamefont
  {Mourik}}, \bibinfo {author} {\bibfnamefont {K.}~\bibnamefont {Zuo}},
  \bibinfo {author} {\bibfnamefont {S.~M.}\ \bibnamefont {Frolov}}, \bibinfo
  {author} {\bibfnamefont {S.~R.}\ \bibnamefont {Plissard}}, \bibinfo {author}
  {\bibfnamefont {E.~P. A.~M.}\ \bibnamefont {Bakkers}}, \ and\ \bibinfo
  {author} {\bibfnamefont {L.~P.}\ \bibnamefont {Kouwenhoven}},\ }\href
  {\doibase 10.1126/science.1222360} {\bibfield  {journal} {\bibinfo  {journal}
  {Science}\ }\textbf {\bibinfo {volume} {336}},\ \bibinfo {pages} {1003}
  (\bibinfo {year} {2012})},\ \Eprint
  {http://arxiv.org/abs/https://science.sciencemag.org/content/336/6084/1003.full.pdf}
  {https://science.sciencemag.org/content/336/6084/1003.full.pdf} \BibitemShut
  {NoStop}%
\bibitem [{\citenamefont {Das}\ \emph {et~al.}(2012)\citenamefont {Das},
  \citenamefont {Ronen}, \citenamefont {Most}, \citenamefont {Oreg},
  \citenamefont {Heiblum},\ and\ \citenamefont {Shtrikman}}]{Das2012}%
  \BibitemOpen
  \bibfield  {author} {\bibinfo {author} {\bibfnamefont {A.}~\bibnamefont
  {Das}}, \bibinfo {author} {\bibfnamefont {Y.}~\bibnamefont {Ronen}}, \bibinfo
  {author} {\bibfnamefont {Y.}~\bibnamefont {Most}}, \bibinfo {author}
  {\bibfnamefont {Y.}~\bibnamefont {Oreg}}, \bibinfo {author} {\bibfnamefont
  {M.}~\bibnamefont {Heiblum}}, \ and\ \bibinfo {author} {\bibfnamefont
  {H.}~\bibnamefont {Shtrikman}},\ }\href {\doibase 10.1038/nphys2479}
  {\bibfield  {journal} {\bibinfo  {journal} {Nat. Phys.}\ }\textbf {\bibinfo
  {volume} {8}},\ \bibinfo {pages} {887} (\bibinfo {year} {2012})}\BibitemShut
  {NoStop}%
\bibitem [{\citenamefont {Deng}\ \emph {et~al.}(2016)\citenamefont {Deng},
  \citenamefont {Vaitiekenas}, \citenamefont {Hansen}, \citenamefont {Danon},
  \citenamefont {Leijnse}, \citenamefont {Flensberg}, \citenamefont {Nyg{\r
  a}rd}, \citenamefont {Krogstrup},\ and\ \citenamefont {Marcus}}]{Deng1557}%
  \BibitemOpen
  \bibfield  {author} {\bibinfo {author} {\bibfnamefont {M.~T.}\ \bibnamefont
  {Deng}}, \bibinfo {author} {\bibfnamefont {S.}~\bibnamefont {Vaitiekenas}},
  \bibinfo {author} {\bibfnamefont {E.~B.}\ \bibnamefont {Hansen}}, \bibinfo
  {author} {\bibfnamefont {J.}~\bibnamefont {Danon}}, \bibinfo {author}
  {\bibfnamefont {M.}~\bibnamefont {Leijnse}}, \bibinfo {author} {\bibfnamefont
  {K.}~\bibnamefont {Flensberg}}, \bibinfo {author} {\bibfnamefont
  {J.}~\bibnamefont {Nyg{\r a}rd}}, \bibinfo {author} {\bibfnamefont
  {P.}~\bibnamefont {Krogstrup}}, \ and\ \bibinfo {author} {\bibfnamefont
  {C.~M.}\ \bibnamefont {Marcus}},\ }\href {\doibase 10.1126/science.aaf3961}
  {\bibfield  {journal} {\bibinfo  {journal} {Science}\ }\textbf {\bibinfo
  {volume} {354}},\ \bibinfo {pages} {1557} (\bibinfo {year} {2016})},\ \Eprint
  {http://arxiv.org/abs/https://science.sciencemag.org/content/354/6319/1557.full.pdf}
  {https://science.sciencemag.org/content/354/6319/1557.full.pdf} \BibitemShut
  {NoStop}%
\bibitem [{\citenamefont {Ugeda}\ \emph {et~al.}(2015)\citenamefont {Ugeda},
  \citenamefont {Bradley}, \citenamefont {Zhang}, \citenamefont {Onishi},
  \citenamefont {Chen}, \citenamefont {Ruan}, \citenamefont
  {Ojeda-Aristizabal}, \citenamefont {Ryu}, \citenamefont {Edmonds},
  \citenamefont {Tsai}, \citenamefont {Riss}, \citenamefont {Mo}, \citenamefont
  {Lee}, \citenamefont {Zettl}, \citenamefont {Hussain}, \citenamefont {Shen},\
  and\ \citenamefont {Crommie}}]{Ugeda2015}%
  \BibitemOpen
  \bibfield  {author} {\bibinfo {author} {\bibfnamefont {M.~M.}\ \bibnamefont
  {Ugeda}}, \bibinfo {author} {\bibfnamefont {A.~J.}\ \bibnamefont {Bradley}},
  \bibinfo {author} {\bibfnamefont {Y.}~\bibnamefont {Zhang}}, \bibinfo
  {author} {\bibfnamefont {S.}~\bibnamefont {Onishi}}, \bibinfo {author}
  {\bibfnamefont {Y.}~\bibnamefont {Chen}}, \bibinfo {author} {\bibfnamefont
  {W.}~\bibnamefont {Ruan}}, \bibinfo {author} {\bibfnamefont {C.}~\bibnamefont
  {Ojeda-Aristizabal}}, \bibinfo {author} {\bibfnamefont {H.}~\bibnamefont
  {Ryu}}, \bibinfo {author} {\bibfnamefont {M.~T.}\ \bibnamefont {Edmonds}},
  \bibinfo {author} {\bibfnamefont {H.-Z.}\ \bibnamefont {Tsai}}, \bibinfo
  {author} {\bibfnamefont {A.}~\bibnamefont {Riss}}, \bibinfo {author}
  {\bibfnamefont {S.-K.}\ \bibnamefont {Mo}}, \bibinfo {author} {\bibfnamefont
  {D.}~\bibnamefont {Lee}}, \bibinfo {author} {\bibfnamefont {A.}~\bibnamefont
  {Zettl}}, \bibinfo {author} {\bibfnamefont {Z.}~\bibnamefont {Hussain}},
  \bibinfo {author} {\bibfnamefont {Z.-X.}\ \bibnamefont {Shen}}, \ and\
  \bibinfo {author} {\bibfnamefont {M.~F.}\ \bibnamefont {Crommie}},\ }\href
  {\doibase 10.1038/nphys3527} {\bibfield  {journal} {\bibinfo  {journal} {Nat.
  Phys.}\ }\textbf {\bibinfo {volume} {12}},\ \bibinfo {pages} {92} (\bibinfo
  {year} {2015})}\BibitemShut {NoStop}%
\bibitem [{\citenamefont {Rokhinson}\ \emph {et~al.}(2012)\citenamefont
  {Rokhinson}, \citenamefont {Liu},\ and\ \citenamefont
  {Furdyna}}]{Rokhinson2012}%
  \BibitemOpen
  \bibfield  {author} {\bibinfo {author} {\bibfnamefont {L.~P.}\ \bibnamefont
  {Rokhinson}}, \bibinfo {author} {\bibfnamefont {X.}~\bibnamefont {Liu}}, \
  and\ \bibinfo {author} {\bibfnamefont {J.~K.}\ \bibnamefont {Furdyna}},\
  }\href {\doibase 10.1038/nphys2429} {\bibfield  {journal} {\bibinfo
  {journal} {Nat. Phys.}\ }\textbf {\bibinfo {volume} {8}},\ \bibinfo {pages}
  {795} (\bibinfo {year} {2012})}\BibitemShut {NoStop}%
\bibitem [{\citenamefont {Cai}\ \emph {et~al.}(2013)\citenamefont {Cai},
  \citenamefont {Lang}, \citenamefont {Chen},\ and\ \citenamefont
  {Wang}}]{PhysRevLett.110.176403}%
  \BibitemOpen
  \bibfield  {author} {\bibinfo {author} {\bibfnamefont {X.}~\bibnamefont
  {Cai}}, \bibinfo {author} {\bibfnamefont {L.-J.}\ \bibnamefont {Lang}},
  \bibinfo {author} {\bibfnamefont {S.}~\bibnamefont {Chen}}, \ and\ \bibinfo
  {author} {\bibfnamefont {Y.}~\bibnamefont {Wang}},\ }\href {\doibase
  10.1103/PhysRevLett.110.176403} {\bibfield  {journal} {\bibinfo  {journal}
  {Phys. Rev. Lett.}\ }\textbf {\bibinfo {volume} {110}},\ \bibinfo {pages}
  {176403} (\bibinfo {year} {2013})}\BibitemShut {NoStop}%
\bibitem [{\citenamefont {DeGottardi}\ \emph {et~al.}(2013)\citenamefont
  {DeGottardi}, \citenamefont {Sen},\ and\ \citenamefont
  {Vishveshwara}}]{PhysRevLett.110.146404}%
  \BibitemOpen
  \bibfield  {author} {\bibinfo {author} {\bibfnamefont {W.}~\bibnamefont
  {DeGottardi}}, \bibinfo {author} {\bibfnamefont {D.}~\bibnamefont {Sen}}, \
  and\ \bibinfo {author} {\bibfnamefont {S.}~\bibnamefont {Vishveshwara}},\
  }\href {\doibase 10.1103/PhysRevLett.110.146404} {\bibfield  {journal}
  {\bibinfo  {journal} {Phys. Rev. Lett.}\ }\textbf {\bibinfo {volume} {110}},\
  \bibinfo {pages} {146404} (\bibinfo {year} {2013})}\BibitemShut {NoStop}%
\bibitem [{\citenamefont {Malard}\ \emph {et~al.}(2016)\citenamefont {Malard},
  \citenamefont {Japaridze},\ and\ \citenamefont
  {Johannesson}}]{PhysRevB.94.115128}%
  \BibitemOpen
  \bibfield  {author} {\bibinfo {author} {\bibfnamefont {M.}~\bibnamefont
  {Malard}}, \bibinfo {author} {\bibfnamefont {G.~I.}\ \bibnamefont
  {Japaridze}}, \ and\ \bibinfo {author} {\bibfnamefont {H.}~\bibnamefont
  {Johannesson}},\ }\href {\doibase 10.1103/PhysRevB.94.115128} {\bibfield
  {journal} {\bibinfo  {journal} {Phys. Rev. B}\ }\textbf {\bibinfo {volume}
  {94}},\ \bibinfo {pages} {115128} (\bibinfo {year} {2016})}\BibitemShut
  {NoStop}%
\bibitem [{\citenamefont {Tezuka}\ and\ \citenamefont
  {Kawakami}(2012)}]{PhysRevB.85.140508}%
  \BibitemOpen
  \bibfield  {author} {\bibinfo {author} {\bibfnamefont {M.}~\bibnamefont
  {Tezuka}}\ and\ \bibinfo {author} {\bibfnamefont {N.}~\bibnamefont
  {Kawakami}},\ }\href {\doibase 10.1103/PhysRevB.85.140508} {\bibfield
  {journal} {\bibinfo  {journal} {Phys. Rev. B}\ }\textbf {\bibinfo {volume}
  {85}},\ \bibinfo {pages} {140508} (\bibinfo {year} {2012})}\BibitemShut
  {NoStop}%
\bibitem [{\citenamefont {Levine}\ \emph {et~al.}(2017)\citenamefont {Levine},
  \citenamefont {Haim},\ and\ \citenamefont {Oreg}}]{PhysRevB.96.165147}%
  \BibitemOpen
  \bibfield  {author} {\bibinfo {author} {\bibfnamefont {Y.}~\bibnamefont
  {Levine}}, \bibinfo {author} {\bibfnamefont {A.}~\bibnamefont {Haim}}, \ and\
  \bibinfo {author} {\bibfnamefont {Y.}~\bibnamefont {Oreg}},\ }\href {\doibase
  10.1103/PhysRevB.96.165147} {\bibfield  {journal} {\bibinfo  {journal} {Phys.
  Rev. B}\ }\textbf {\bibinfo {volume} {96}},\ \bibinfo {pages} {165147}
  (\bibinfo {year} {2017})}\BibitemShut {NoStop}%
\bibitem [{\citenamefont {B{\o}ttcher}\ \emph {et~al.}(2018)\citenamefont
  {B{\o}ttcher}, \citenamefont {Nichele}, \citenamefont {Kjaergaard},
  \citenamefont {Suominen}, \citenamefont {Shabani}, \citenamefont
  {Palmstr{\o}m},\ and\ \citenamefont {Marcus}}]{Bttcher2018}%
  \BibitemOpen
  \bibfield  {author} {\bibinfo {author} {\bibfnamefont {C.~G.~L.}\
  \bibnamefont {B{\o}ttcher}}, \bibinfo {author} {\bibfnamefont
  {F.}~\bibnamefont {Nichele}}, \bibinfo {author} {\bibfnamefont
  {M.}~\bibnamefont {Kjaergaard}}, \bibinfo {author} {\bibfnamefont {H.~J.}\
  \bibnamefont {Suominen}}, \bibinfo {author} {\bibfnamefont {J.}~\bibnamefont
  {Shabani}}, \bibinfo {author} {\bibfnamefont {C.~J.}\ \bibnamefont
  {Palmstr{\o}m}}, \ and\ \bibinfo {author} {\bibfnamefont {C.~M.}\
  \bibnamefont {Marcus}},\ }\href {\doibase 10.1038/s41567-018-0259-9}
  {\bibfield  {journal} {\bibinfo  {journal} {Nat. Phys.}\ }\textbf {\bibinfo
  {volume} {14}},\ \bibinfo {pages} {1138} (\bibinfo {year}
  {2018})}\BibitemShut {NoStop}%
\bibitem [{\citenamefont {Laeven}\ \emph {et~al.}(2020)\citenamefont {Laeven},
  \citenamefont {Nijholt}, \citenamefont {Wimmer},\ and\ \citenamefont
  {Akhmerov}}]{PhysRevLett.125.086802}%
  \BibitemOpen
  \bibfield  {author} {\bibinfo {author} {\bibfnamefont {T.}~\bibnamefont
  {Laeven}}, \bibinfo {author} {\bibfnamefont {B.}~\bibnamefont {Nijholt}},
  \bibinfo {author} {\bibfnamefont {M.}~\bibnamefont {Wimmer}}, \ and\ \bibinfo
  {author} {\bibfnamefont {A.~R.}\ \bibnamefont {Akhmerov}},\ }\href {\doibase
  10.1103/PhysRevLett.125.086802} {\bibfield  {journal} {\bibinfo  {journal}
  {Phys. Rev. Lett.}\ }\textbf {\bibinfo {volume} {125}},\ \bibinfo {pages}
  {086802} (\bibinfo {year} {2020})}\BibitemShut {NoStop}%
\bibitem [{\citenamefont {Harper}(1955)}]{Harper1955}%
  \BibitemOpen
  \bibfield  {author} {\bibinfo {author} {\bibfnamefont {P.~G.}\ \bibnamefont
  {Harper}},\ }\href {\doibase 10.1088/0370-1298/68/10/304} {\bibfield
  {journal} {\bibinfo  {journal} {Proc. Phys. Soc. A}\ }\textbf {\bibinfo
  {volume} {68}},\ \bibinfo {pages} {874} (\bibinfo {year} {1955})}\BibitemShut
  {NoStop}%
\bibitem [{\citenamefont {Lutchyn}\ \emph {et~al.}(2010)\citenamefont
  {Lutchyn}, \citenamefont {Sau},\ and\ \citenamefont
  {Das~Sarma}}]{PhysRevLett.105.077001}%
  \BibitemOpen
  \bibfield  {author} {\bibinfo {author} {\bibfnamefont {R.~M.}\ \bibnamefont
  {Lutchyn}}, \bibinfo {author} {\bibfnamefont {J.~D.}\ \bibnamefont {Sau}}, \
  and\ \bibinfo {author} {\bibfnamefont {S.}~\bibnamefont {Das~Sarma}},\ }\href
  {\doibase 10.1103/PhysRevLett.105.077001} {\bibfield  {journal} {\bibinfo
  {journal} {Phys. Rev. Lett.}\ }\textbf {\bibinfo {volume} {105}},\ \bibinfo
  {pages} {077001} (\bibinfo {year} {2010})}\BibitemShut {NoStop}%
\bibitem [{\citenamefont {Thouless}\ \emph {et~al.}(1982)\citenamefont
  {Thouless}, \citenamefont {Kohmoto}, \citenamefont {Nightingale},\ and\
  \citenamefont {den Nijs}}]{PhysRevLett.49.405}%
  \BibitemOpen
  \bibfield  {author} {\bibinfo {author} {\bibfnamefont {D.~J.}\ \bibnamefont
  {Thouless}}, \bibinfo {author} {\bibfnamefont {M.}~\bibnamefont {Kohmoto}},
  \bibinfo {author} {\bibfnamefont {M.~P.}\ \bibnamefont {Nightingale}}, \ and\
  \bibinfo {author} {\bibfnamefont {M.}~\bibnamefont {den Nijs}},\ }\href
  {\doibase 10.1103/PhysRevLett.49.405} {\bibfield  {journal} {\bibinfo
  {journal} {Phys. Rev. Lett.}\ }\textbf {\bibinfo {volume} {49}},\ \bibinfo
  {pages} {405} (\bibinfo {year} {1982})}\BibitemShut {NoStop}%
\bibitem [{\citenamefont {Bistritzer}\ and\ \citenamefont
  {MacDonald}(2011)}]{Bistritzer2011}%
  \BibitemOpen
  \bibfield  {author} {\bibinfo {author} {\bibfnamefont {R.}~\bibnamefont
  {Bistritzer}}\ and\ \bibinfo {author} {\bibfnamefont {A.~H.}\ \bibnamefont
  {MacDonald}},\ }\href {\doibase 10.1073/pnas.1108174108} {\bibfield
  {journal} {\bibinfo  {journal} {Proc. Nat. Acad. Sci.}\ }\textbf {\bibinfo
  {volume} {108}},\ \bibinfo {pages} {12233} (\bibinfo {year}
  {2011})}\BibitemShut {NoStop}%
\bibitem [{\citenamefont {Su\'arez~Morell}\ \emph {et~al.}(2010)\citenamefont
  {Su\'arez~Morell}, \citenamefont {Correa}, \citenamefont {Vargas},
  \citenamefont {Pacheco},\ and\ \citenamefont
  {Barticevic}}]{PhysRevB.82.121407}%
  \BibitemOpen
  \bibfield  {author} {\bibinfo {author} {\bibfnamefont {E.}~\bibnamefont
  {Su\'arez~Morell}}, \bibinfo {author} {\bibfnamefont {J.~D.}\ \bibnamefont
  {Correa}}, \bibinfo {author} {\bibfnamefont {P.}~\bibnamefont {Vargas}},
  \bibinfo {author} {\bibfnamefont {M.}~\bibnamefont {Pacheco}}, \ and\
  \bibinfo {author} {\bibfnamefont {Z.}~\bibnamefont {Barticevic}},\ }\href
  {\doibase 10.1103/PhysRevB.82.121407} {\bibfield  {journal} {\bibinfo
  {journal} {Phys. Rev. B}\ }\textbf {\bibinfo {volume} {82}},\ \bibinfo
  {pages} {121407} (\bibinfo {year} {2010})}\BibitemShut {NoStop}%
\bibitem [{\citenamefont {Algra}\ \emph {et~al.}(2008)\citenamefont {Algra},
  \citenamefont {Verheijen}, \citenamefont {Borgstr\"{o}m}, \citenamefont
  {Feiner}, \citenamefont {Immink}, \citenamefont {van Enckevort},
  \citenamefont {Vlieg},\ and\ \citenamefont {Bakkers}}]{Algra2008}%
  \BibitemOpen
  \bibfield  {author} {\bibinfo {author} {\bibfnamefont {R.~E.}\ \bibnamefont
  {Algra}}, \bibinfo {author} {\bibfnamefont {M.~A.}\ \bibnamefont
  {Verheijen}}, \bibinfo {author} {\bibfnamefont {M.~T.}\ \bibnamefont
  {Borgstr\"{o}m}}, \bibinfo {author} {\bibfnamefont {L.-F.}\ \bibnamefont
  {Feiner}}, \bibinfo {author} {\bibfnamefont {G.}~\bibnamefont {Immink}},
  \bibinfo {author} {\bibfnamefont {W.~J.~P.}\ \bibnamefont {van Enckevort}},
  \bibinfo {author} {\bibfnamefont {E.}~\bibnamefont {Vlieg}}, \ and\ \bibinfo
  {author} {\bibfnamefont {E.~P. A.~M.}\ \bibnamefont {Bakkers}},\ }\href
  {\doibase 10.1038/nature07570} {\bibfield  {journal} {\bibinfo  {journal}
  {Nature}\ }\textbf {\bibinfo {volume} {456}},\ \bibinfo {pages} {369}
  (\bibinfo {year} {2008})}\BibitemShut {NoStop}%
\bibitem [{\citenamefont {Zhang}\ \emph {et~al.}(2019)\citenamefont {Zhang},
  \citenamefont {Takiguchi}, \citenamefont {Tateno}, \citenamefont {Tawara},
  \citenamefont {Notomi},\ and\ \citenamefont {Gotoh}}]{Zhang2019}%
  \BibitemOpen
  \bibfield  {author} {\bibinfo {author} {\bibfnamefont {G.}~\bibnamefont
  {Zhang}}, \bibinfo {author} {\bibfnamefont {M.}~\bibnamefont {Takiguchi}},
  \bibinfo {author} {\bibfnamefont {K.}~\bibnamefont {Tateno}}, \bibinfo
  {author} {\bibfnamefont {T.}~\bibnamefont {Tawara}}, \bibinfo {author}
  {\bibfnamefont {M.}~\bibnamefont {Notomi}}, \ and\ \bibinfo {author}
  {\bibfnamefont {H.}~\bibnamefont {Gotoh}},\ }\href {\doibase
  10.1126/sciadv.aat8896} {\bibfield  {journal} {\bibinfo  {journal} {Sci.
  Adv.}\ }\textbf {\bibinfo {volume} {5}},\ \bibinfo {pages} {eaat8896}
  (\bibinfo {year} {2019})}\BibitemShut {NoStop}%
\bibitem [{\citenamefont {Kezilebieke}\ \emph {et~al.}(2021)\citenamefont
  {Kezilebieke}, \citenamefont {Silveira}, \citenamefont {Huda}, \citenamefont
  {Va\v{n}o}, \citenamefont {Aapro}, \citenamefont {Ganguli}, \citenamefont
  {Lahtinen}, \citenamefont {Mansell}, \citenamefont {van Dijken},
  \citenamefont {Foster},\ and\ \citenamefont {Liljeroth}}]{Kezilebieke2020}%
  \BibitemOpen
  \bibfield  {author} {\bibinfo {author} {\bibfnamefont {S.}~\bibnamefont
  {Kezilebieke}}, \bibinfo {author} {\bibfnamefont {O.~J.}\ \bibnamefont
  {Silveira}}, \bibinfo {author} {\bibfnamefont {M.~N.}\ \bibnamefont {Huda}},
  \bibinfo {author} {\bibfnamefont {V.}~\bibnamefont {Va\v{n}o}}, \bibinfo
  {author} {\bibfnamefont {M.}~\bibnamefont {Aapro}}, \bibinfo {author}
  {\bibfnamefont {S.~C.}\ \bibnamefont {Ganguli}}, \bibinfo {author}
  {\bibfnamefont {J.}~\bibnamefont {Lahtinen}}, \bibinfo {author}
  {\bibfnamefont {R.}~\bibnamefont {Mansell}}, \bibinfo {author} {\bibfnamefont
  {S.}~\bibnamefont {van Dijken}}, \bibinfo {author} {\bibfnamefont {A.~S.}\
  \bibnamefont {Foster}}, \ and\ \bibinfo {author} {\bibfnamefont
  {P.}~\bibnamefont {Liljeroth}},\ }\href {\doibase
  https://doi.org/10.1002/adma.202006850} {\bibfield  {journal} {\bibinfo
  {journal} {Adv. Mater.}\ }\textbf {\bibinfo {volume} {33}},\ \bibinfo {pages}
  {2006850} (\bibinfo {year} {2021})}\BibitemShut {NoStop}%
\bibitem [{\citenamefont {Chen}\ \emph {et~al.}(2019)\citenamefont {Chen},
  \citenamefont {Sun}, \citenamefont {Wang}, \citenamefont {Gu}, \citenamefont
  {Xu}, \citenamefont {Wu},\ and\ \citenamefont {Gao}}]{Chen2019}%
  \BibitemOpen
  \bibfield  {author} {\bibinfo {author} {\bibfnamefont {W.}~\bibnamefont
  {Chen}}, \bibinfo {author} {\bibfnamefont {Z.}~\bibnamefont {Sun}}, \bibinfo
  {author} {\bibfnamefont {Z.}~\bibnamefont {Wang}}, \bibinfo {author}
  {\bibfnamefont {L.}~\bibnamefont {Gu}}, \bibinfo {author} {\bibfnamefont
  {X.}~\bibnamefont {Xu}}, \bibinfo {author} {\bibfnamefont {S.}~\bibnamefont
  {Wu}}, \ and\ \bibinfo {author} {\bibfnamefont {C.}~\bibnamefont {Gao}},\
  }\href {\doibase 10.1126/science.aav1937} {\bibfield  {journal} {\bibinfo
  {journal} {Science}\ }\textbf {\bibinfo {volume} {366}},\ \bibinfo {pages}
  {983} (\bibinfo {year} {2019})}\BibitemShut {NoStop}%
\bibitem [{\citenamefont {Sivadas}\ \emph {et~al.}(2018)\citenamefont
  {Sivadas}, \citenamefont {Okamoto}, \citenamefont {Xu}, \citenamefont
  {Fennie},\ and\ \citenamefont {Xiao}}]{Sivadas2018}%
  \BibitemOpen
  \bibfield  {author} {\bibinfo {author} {\bibfnamefont {N.}~\bibnamefont
  {Sivadas}}, \bibinfo {author} {\bibfnamefont {S.}~\bibnamefont {Okamoto}},
  \bibinfo {author} {\bibfnamefont {X.}~\bibnamefont {Xu}}, \bibinfo {author}
  {\bibfnamefont {C.~J.}\ \bibnamefont {Fennie}}, \ and\ \bibinfo {author}
  {\bibfnamefont {D.}~\bibnamefont {Xiao}},\ }\href {\doibase
  10.1021/acs.nanolett.8b03321} {\bibfield  {journal} {\bibinfo  {journal}
  {Nano Lett.}\ }\textbf {\bibinfo {volume} {18}},\ \bibinfo {pages} {7658}
  (\bibinfo {year} {2018})}\BibitemShut {NoStop}%
\bibitem [{\citenamefont {Song}\ \emph {et~al.}(2019)\citenamefont {Song},
  \citenamefont {Fei}, \citenamefont {Yankowitz}, \citenamefont {Lin},
  \citenamefont {Jiang}, \citenamefont {Hwangbo}, \citenamefont {Zhang},
  \citenamefont {Sun}, \citenamefont {Taniguchi}, \citenamefont {Watanabe},
  \citenamefont {McGuire}, \citenamefont {Graf}, \citenamefont {Cao},
  \citenamefont {Chu}, \citenamefont {Cobden}, \citenamefont {Dean},
  \citenamefont {Xiao},\ and\ \citenamefont {Xu}}]{Song2019}%
  \BibitemOpen
  \bibfield  {author} {\bibinfo {author} {\bibfnamefont {T.}~\bibnamefont
  {Song}}, \bibinfo {author} {\bibfnamefont {Z.}~\bibnamefont {Fei}}, \bibinfo
  {author} {\bibfnamefont {M.}~\bibnamefont {Yankowitz}}, \bibinfo {author}
  {\bibfnamefont {Z.}~\bibnamefont {Lin}}, \bibinfo {author} {\bibfnamefont
  {Q.}~\bibnamefont {Jiang}}, \bibinfo {author} {\bibfnamefont
  {K.}~\bibnamefont {Hwangbo}}, \bibinfo {author} {\bibfnamefont
  {Q.}~\bibnamefont {Zhang}}, \bibinfo {author} {\bibfnamefont
  {B.}~\bibnamefont {Sun}}, \bibinfo {author} {\bibfnamefont {T.}~\bibnamefont
  {Taniguchi}}, \bibinfo {author} {\bibfnamefont {K.}~\bibnamefont {Watanabe}},
  \bibinfo {author} {\bibfnamefont {M.~A.}\ \bibnamefont {McGuire}}, \bibinfo
  {author} {\bibfnamefont {D.}~\bibnamefont {Graf}}, \bibinfo {author}
  {\bibfnamefont {T.}~\bibnamefont {Cao}}, \bibinfo {author} {\bibfnamefont
  {J.-H.}\ \bibnamefont {Chu}}, \bibinfo {author} {\bibfnamefont {D.~H.}\
  \bibnamefont {Cobden}}, \bibinfo {author} {\bibfnamefont {C.~R.}\
  \bibnamefont {Dean}}, \bibinfo {author} {\bibfnamefont {D.}~\bibnamefont
  {Xiao}}, \ and\ \bibinfo {author} {\bibfnamefont {X.}~\bibnamefont {Xu}},\
  }\href {\doibase 10.1038/s41563-019-0505-2} {\bibfield  {journal} {\bibinfo
  {journal} {Nat. Mater.}\ }\textbf {\bibinfo {volume} {18}},\ \bibinfo {pages}
  {1298} (\bibinfo {year} {2019})}\BibitemShut {NoStop}%
\bibitem [{\citenamefont {Wickramaratne}\ \emph {et~al.}(2020)\citenamefont
  {Wickramaratne}, \citenamefont {Khmelevskyi}, \citenamefont {Agterberg},\
  and\ \citenamefont {Mazin}}]{PhysRevX.10.041003}%
  \BibitemOpen
  \bibfield  {author} {\bibinfo {author} {\bibfnamefont {D.}~\bibnamefont
  {Wickramaratne}}, \bibinfo {author} {\bibfnamefont {S.}~\bibnamefont
  {Khmelevskyi}}, \bibinfo {author} {\bibfnamefont {D.~F.}\ \bibnamefont
  {Agterberg}}, \ and\ \bibinfo {author} {\bibfnamefont {I.~I.}\ \bibnamefont
  {Mazin}},\ }\href {\doibase 10.1103/PhysRevX.10.041003} {\bibfield  {journal}
  {\bibinfo  {journal} {Phys. Rev. X}\ }\textbf {\bibinfo {volume} {10}},\
  \bibinfo {pages} {041003} (\bibinfo {year} {2020})}\BibitemShut {NoStop}%
\bibitem [{\citenamefont {Naik}\ and\ \citenamefont
  {Jain}(2018)}]{PhysRevLett.121.266401}%
  \BibitemOpen
  \bibfield  {author} {\bibinfo {author} {\bibfnamefont {M.~H.}\ \bibnamefont
  {Naik}}\ and\ \bibinfo {author} {\bibfnamefont {M.}~\bibnamefont {Jain}},\
  }\href {\doibase 10.1103/PhysRevLett.121.266401} {\bibfield  {journal}
  {\bibinfo  {journal} {Phys. Rev. Lett.}\ }\textbf {\bibinfo {volume} {121}},\
  \bibinfo {pages} {266401} (\bibinfo {year} {2018})}\BibitemShut {NoStop}%
\bibitem [{\citenamefont {Enaldiev}\ \emph {et~al.}(2020)\citenamefont
  {Enaldiev}, \citenamefont {Z\'olyomi}, \citenamefont {Yelgel}, \citenamefont
  {Magorrian},\ and\ \citenamefont {Fal'ko}}]{PhysRevLett.124.206101}%
  \BibitemOpen
  \bibfield  {author} {\bibinfo {author} {\bibfnamefont {V.~V.}\ \bibnamefont
  {Enaldiev}}, \bibinfo {author} {\bibfnamefont {V.}~\bibnamefont {Z\'olyomi}},
  \bibinfo {author} {\bibfnamefont {C.}~\bibnamefont {Yelgel}}, \bibinfo
  {author} {\bibfnamefont {S.~J.}\ \bibnamefont {Magorrian}}, \ and\ \bibinfo
  {author} {\bibfnamefont {V.~I.}\ \bibnamefont {Fal'ko}},\ }\href {\doibase
  10.1103/PhysRevLett.124.206101} {\bibfield  {journal} {\bibinfo  {journal}
  {Phys. Rev. Lett.}\ }\textbf {\bibinfo {volume} {124}},\ \bibinfo {pages}
  {206101} (\bibinfo {year} {2020})}\BibitemShut {NoStop}%
\bibitem [{\citenamefont {Zheng}\ \emph {et~al.}(2018)\citenamefont {Zheng},
  \citenamefont {Zhou}, \citenamefont {Liu},\ and\ \citenamefont
  {Feng}}]{PhysRevB.97.081101}%
  \BibitemOpen
  \bibfield  {author} {\bibinfo {author} {\bibfnamefont {F.}~\bibnamefont
  {Zheng}}, \bibinfo {author} {\bibfnamefont {Z.}~\bibnamefont {Zhou}},
  \bibinfo {author} {\bibfnamefont {X.}~\bibnamefont {Liu}}, \ and\ \bibinfo
  {author} {\bibfnamefont {J.}~\bibnamefont {Feng}},\ }\href {\doibase
  10.1103/PhysRevB.97.081101} {\bibfield  {journal} {\bibinfo  {journal} {Phys.
  Rev. B}\ }\textbf {\bibinfo {volume} {97}},\ \bibinfo {pages} {081101}
  (\bibinfo {year} {2018})}\BibitemShut {NoStop}%
\bibitem [{\citenamefont {Guster}\ \emph {et~al.}(2019)\citenamefont {Guster},
  \citenamefont {Rubio-Verd{\'{u}}}, \citenamefont {Robles}, \citenamefont
  {Zald{\'{\i}}var}, \citenamefont {Dreher}, \citenamefont {Pruneda},
  \citenamefont {Silva-Guill{\'{e}}n}, \citenamefont {Choi}, \citenamefont
  {Pascual}, \citenamefont {Ugeda}, \citenamefont {Ordej{\'{o}}n},\ and\
  \citenamefont {Canadell}}]{Guster2019}%
  \BibitemOpen
  \bibfield  {author} {\bibinfo {author} {\bibfnamefont {B.}~\bibnamefont
  {Guster}}, \bibinfo {author} {\bibfnamefont {C.}~\bibnamefont
  {Rubio-Verd{\'{u}}}}, \bibinfo {author} {\bibfnamefont {R.}~\bibnamefont
  {Robles}}, \bibinfo {author} {\bibfnamefont {J.}~\bibnamefont
  {Zald{\'{\i}}var}}, \bibinfo {author} {\bibfnamefont {P.}~\bibnamefont
  {Dreher}}, \bibinfo {author} {\bibfnamefont {M.}~\bibnamefont {Pruneda}},
  \bibinfo {author} {\bibfnamefont {J.~{\'{A}}.}\ \bibnamefont
  {Silva-Guill{\'{e}}n}}, \bibinfo {author} {\bibfnamefont {D.-J.}\
  \bibnamefont {Choi}}, \bibinfo {author} {\bibfnamefont {J.~I.}\ \bibnamefont
  {Pascual}}, \bibinfo {author} {\bibfnamefont {M.~M.}\ \bibnamefont {Ugeda}},
  \bibinfo {author} {\bibfnamefont {P.}~\bibnamefont {Ordej{\'{o}}n}}, \ and\
  \bibinfo {author} {\bibfnamefont {E.}~\bibnamefont {Canadell}},\ }\href
  {\doibase 10.1021/acs.nanolett.9b00268} {\bibfield  {journal} {\bibinfo
  {journal} {Nano Lett.}\ }\textbf {\bibinfo {volume} {19}},\ \bibinfo {pages}
  {3027} (\bibinfo {year} {2019})}\BibitemShut {NoStop}%
\end{thebibliography}%

\end{document}